\documentclass[showpacs,amsmath,amssymb,twocolumn,prx,superscriptaddress,10pt,aps,nofootinbib]{revtex4-2}
\pdfoutput=1

\usepackage[utf8]{inputenc} 

\usepackage{tikz}
\usetikzlibrary{quantikz2}

\usepackage{siunitx}
\usepackage{amsmath,amssymb,amsthm,amsfonts}
\usepackage{subfigure, epsfig}
\usepackage{braket}
\usepackage{bm}
\usepackage{enumerate}
\usepackage[dvipsnames]{xcolor}
\usepackage{color}
\usepackage{fancybox}
\usepackage{graphicx}
\usepackage[skins]{tcolorbox}
\usepackage{multirow}
\usepackage{mathtools}
\usepackage{makecell}
\usepackage{enumitem}

\usepackage[colorlinks]{hyperref}
\hypersetup{
	colorlinks=true,
	citecolor = {blue},
    linkcolor=blue,
    filecolor=magenta,      
    urlcolor=cyan,
}
\usepackage{cleveref}

\newcommand{\thm}[1]{\hyperref[thm:#1]{Theorem~\ref*{thm:#1}}}
\newcommand{\defn}[1]{\hyperref[defn:#1]{Definition~\ref*{defn:#1}}}
\newcommand{\lem}[1]{\hyperref[lem:#1]{Lemma~\ref*{lem:#1}}}
\newcommand{\prop}[1]{\hyperref[prop:#1]{Proposition~\ref*{prop:#1}}}
\newcommand{\fig}[1]{\hyperref[Fig:#1]{Figure~\ref*{Fig:#1}}}
\newcommand{\tab}[1]{\hyperref[tab:#1]{Table~\ref*{tab:#1}}}
\renewcommand{\sec}[1]{\hyperref[Sec:#1]{Section~\ref*{Sec:#1}}}
\newcommand{\append}[1]{\hyperref[App:#1]{Appendix~\ref*{App:#1}}}
\newcommand{\cor}[1]{\hyperref[cor:#1]{Corollary~\ref*{cor:#1}}}
\newcommand{\obs}[1]{\hyperref[obs:#1]{Observation~\ref*{obs:#1}}}

\newcommand{\vertiii}[1]{{\left\vert\kern-0.25ex\left\vert\kern-0.25ex\left\vert #1
		\right\vert\kern-0.25ex\right\vert\kern-0.25ex\right\vert}}

\newcommand{\E}{\mathbb{E}}

\allowdisplaybreaks

\usepackage{titlesec}
\begin{document}

\title{Suppressing errors in analog logical rotation gates via balanced fusion}

\author{Sam McArdle}
\email{sam.mcardle.science@gmail.com}
\thanks{Current affiliation: NVIDIA Corporation, Santa Clara, CA, USA}
\affiliation{AWS Center for Quantum Computing, Pasadena, CA 91125, USA}

\author{Alexander M. Dalzell}
\affiliation{AWS Center for Quantum Computing, Pasadena, CA 91125, USA}

\author{Fernando G.S.L. Brand\~ao}
\affiliation{AWS Center for Quantum Computing, Pasadena, CA 91125, USA}

\affiliation{Institute for Quantum Information and Matter, \\
California Institute of Technology, Pasadena, California 91125, USA}

\begin{abstract}
There have been a number of recent proposals to use analog logical rotations in early fault-tolerant quantum algorithms. Existing proposals implement a logical rotation by angle $\phi$, with error $\mathcal{O}(p\phi)$, where $p$ is the physical error rate. While this is not fault-tolerant, if $\phi$ is sufficiently small the logical error rate can be suppressed. In this work, we introduce and analyze the `balanced fusion' technique for improving the error scaling of analog logical rotations to $\mathcal{O}(p \phi^{1.5})$, without having to resort to fallback synthesis. Balanced fusion is enabled by an improved analysis of the accumulation of coherent error terms in analog logical rotations. Our techniques improve the viability of analog logical rotations for small rotation angles as an alternative to cultivation-powered rotation synthesis.
\end{abstract}

\maketitle

In NISQ computation on physical qubits, we typically treat physical single-qubit gates as being low-cost operations, because they have much lower error rates than physical two-qubit gates. In fault-tolerant computation on logical qubits, this accounting is reversed. As no quantum error correcting code can implement $R_z(\phi) := \exp(i\phi Z) = \mathrm{Diag}[e^{i\phi}, e^{-i\phi}]$ gates transversally (for arbitrary $\phi$)~\cite{eastin2009restrictions}, we typically implement these gates by 1) Decomposing $R_z(\phi)$ into a number of $T$ + Clifford gates (synthesis)~\cite{dawson2005solovay}, 2) Replacing each $T$ gate by an equivalent teleportation gadget which consumes a $T$ magic state, and 3) Preparing the required $T$ states through specialized subroutines, such as magic state distillation~\cite{litinski2019magic} or cultivation~\cite{gidney2024magic}. The spacetime volume to implement a logical $R_z(\phi)$ gate to error $10^{-8}$ can be estimated at approximately $10^7$ qubitrounds\footnote{A qubitround is a unit of spacetime volume that accounts for one qubit being used for a single round of surface code syndrome extraction (4 two-qubit gates, one measurement, one reset). Magic state cultivation can produce $T$ gates with error $10^{-9}$ using $10^5$ qubitrounds~\cite{gidney2024magic}. It is possible to synthesize an $R_z$ gate using $3\log_2(1/\epsilon_s)$ $T$ gates and no ancilla qubits~\cite{ross2016optimal}, where $\epsilon_s$ is the synthesis error in the rotation.} for a standard surface code architecture.

There have been several recent proposals~\cite{choi2023rotations,toshio2025PartiallyFT,zhang2025low,huang2025robust,yoshioka2025transversal,toshio2026starmutation} to implement logical $R_z(\phi)$ gates with lower cost. We will collectively refer to these proposals as implementing `analog logical rotations'. In particular, these proposals target $\phi \ll 1$, and can prepare magic states approximating $\ket{M(\phi)} = \frac{1}{\sqrt{2}}( e^{i\phi}\ket{0} + e^{-i\phi} \ket{1})$ with infidelity $\mathcal{O}(p \phi^2)$ and trace distance $\mathcal{O}(p \phi)$, where $p$ is the physical error rate. This scaling with $p$ means that the proposals are not fault-tolerant. However, under idealized conditions the logical error rate may be suppressed for a sufficiently small target angle. Various techniques have been proposed to mitigate the discrepancy between infidelity and trace distance~\cite{toshio2025PartiallyFT,toshio2026starmutation,zhang2025low}. In App.~\ref{App:MitigateCoherentErrors}, we provide a novel analysis based on the Mixing Lemma~\cite{campbell2017Mixing, hastings2016turning}, showing that problematic off-diagonal errors are naturally controlled in protocols which consume analog logical rotation magic states. By removing the effect of the off-diagonal terms, the trace distance and infidelity coincide to leading order. 
Prior works have proposed Trotterized dynamics of physical systems as a target application for analog logical rotations~\cite{toshio2025PartiallyFT, akahoshi2024compilation, chung2026partiallyMegaQuop}. This is because the angles in a first-order Trotter decomposition are inversely proportional to the simulation time $\mathcal{T}$.

A drawback of the analog logical rotations method is that consuming the magic state $\ket{M(\phi)}$ via teleportation fails with probability $1/2$. This necessitates an equally failure-prone correction using a magic state $\ket{M(2\phi)}$, and so on. Assuming an error on the magic states of $\mathcal{O}(p\phi^2)$ and that error accumulates according to the triangle inequality, after $k$ steps of the baseline repeat-until-success (RUS) process, the average error of the channel is proportional to~\cite{toshio2025PartiallyFT,zeng2025errorstructuretailoredearlyfaulttolerantquantum}:
\begin{equation}\label{Eq:RUS_acc}
    \sum_{i=1}^k \underbrace{\frac{1}{2^{i-1}}}_{\substack{\text{prob of} \\ \text{reaching trial $i$}}} \underbrace{(2^{i-1} \phi )^2 p}_{\substack{\text{error in}\\\text{trial $i$}}}  <   2^k \phi^2 p .
\end{equation}

Consider the special case where $\phi = \pi/2^{k+2}$. In this case, the RUS process terminates (with a Clifford $S$ gate) after $k$ failures. Because $2^{k+1} = \pi/2\phi$, the resulting error in the RUS process is $\mathcal{O}(p\phi)$ --- i.e., the expected error in the gate is only suppressed linearly in the target angle, rather than quadratically. Returning to the Trotterized dynamics example above, a linear error scaling with $\phi$ implies that the circuit is limited by error accumulation to a maximum total simulation time $\mathcal{T} = \mathcal{O}(1)$.

In this work, we develop and analyze a method to mitigate the error accumulation detailed above. We refer to our technique as `balanced fusion', because of its approach for preparing resource states. While the baseline RUS method described above prepares and consumes resource states $\ket{M(2^m\phi)}$ (with error $\mathcal{O}(2^{2m} \phi^2p)$), our method fuses resource states of the form $\ket{M(2^{m/2}\phi)}$, to prepare $\ket{M(2^m\phi)}$ with expected error $\mathcal{O}(2^{1.5m} \phi^2p)$. The resulting error in the $R_z(\phi)$ gate is improved to $\mathcal{O}(p \phi^{1.5})$. We also analyze combining balanced fusion with fallback to Clifford + $T$ synthesis, if the number of RUS trials exceeds a preset threshold $\kappa$, although we show that resorting to fallback is not necessary to achieve the $\mathcal{O}(p \phi^{1.5})$ error scaling. We note that the use of fallback with the baseline RUS method was recently independently proposed by Ref.~\cite{toshio2026starmutation}. 

In Sec.~\ref{Sec:BF}, we explain the balanced fusion method in more detail. In Sec.~\ref{Sec:Apps}, we discuss applications of our technique. In App.~\ref{App:Overview}, we provide a basic overview of the analog logical rotations approach. In App.~\ref{App:MitigateCoherentErrors}, we give a detailed analysis of the natural decoherence of off-diagonal terms, which ensures that the infidelity and trace distance effectively coincide for RUS and balanced fusion of analog logical rotations. Finally, in App.~\ref{App:Comparison}, we provide a summary of, and careful comparison to, prior work.

\section{Balanced fusion}\label{Sec:BF}

\begin{figure*}
    \centering
%
%
\begin{tikzpicture}[
    >=stealth,
    state/.style={draw, circle, minimum size=2.4mm, inner sep=0pt,
                  fill=blue!12},
    discarded/.style={draw, circle, minimum size=2.4mm, inner sep=0pt,
                      fill=gray!25, draw=gray!60!black},
    rus/.style={draw, rounded corners=2pt, fill=yellow!45,
                minimum width=8mm, minimum height=5.5mm, font=\scriptsize},
    succ/.style={very thin, draw=green!45!black},
    fail/.style={very thin, dashed, draw=red!75!black},
    qubit/.style={very thick, ->, draw=black!70},
    trial/.style={font=\bfseries\normalsize},
    lvl/.style={font=\fontsize{6.25}{7.5}\selectfont, anchor=east, text=black!60},
    olbl/.style={font=\small, anchor=west, text=black!70},
    leafl/.style={font=\tiny, align=center},
]

\def\ystep{0.7125}
\def\rusy{3.45}

\foreach \L/\lab in {0/{$\ket{M(\phi)}$},
                     1/{$\ket{M(2\phi)}$},
                     2/{$\ket{M(4\phi)}$},
                     3/{$\ket{M(8\phi)}$},
                     4/{$\ket{M(16\phi)}$}} {
    \draw[densely dotted, draw=black!20]
        (-0.4, {\L*\ystep}) -- (15.0, {\L*\ystep});
    \node[lvl] at (-0.45, {\L*\ystep}) {\lab};
}

\begin{scope}[xshift=0cm]
    \node[state] (T1a) at (0, 0) {};
    \node[rus] (T1r) at (0, \rusy) {RUS$_1$};
    \draw[succ] (T1a) -- (T1r);
    \node[trial] at (0, -0.9) {$i=1$};
\end{scope}

\begin{scope}[xshift=1.6cm]
    \foreach \i in {0,1,2,3} {
        \node[state] (T2a\i) at ({0.32*\i}, 0) {};
    }
    \node[state] (T2s) at (0.16, \ystep) {};
    \draw[succ] (T2a0) -- (T2s);
    \draw[succ] (T2a1) -- (T2s);
    \node[discarded] (T2d) at (0.80, \ystep) {};
    \draw[fail] (T2a2) -- (T2d);
    \draw[fail] (T2a3) -- (T2d);
    \node[rus] (T2r) at (0.16, \rusy) {RUS$_2$};
    \draw[succ] (T2s) -- (T2r);
    \node[trial] at (0.48, -0.9) {$i=2$};
\end{scope}

\begin{scope}[xshift=3.6cm]
    \foreach \i in {0,1,2,3} {
        \node[state] (T3a\i) at ({0.32*\i}, \ystep) {};
    }
    \node[state] (T3s) at (0.16, {2*\ystep}) {};
    \draw[succ] (T3a0) -- (T3s);
    \draw[succ] (T3a1) -- (T3s);
    \node[discarded] (T3d) at (0.80, {2*\ystep}) {};
    \draw[fail] (T3a2) -- (T3d);
    \draw[fail] (T3a3) -- (T3d);
    \node[rus] (T3r) at (0.16, \rusy) {RUS$_3$};
    \draw[succ] (T3s) -- (T3r);
    \node[trial] at (0.48, -0.9) {$i=3$};
\end{scope}

\begin{scope}[xshift=5.4cm]
    \foreach \i in {0,...,15} {
        \node[state] (T4a\i) at ({0.28*\i}, \ystep) {};
    }
    \node[state] (T4b0) at (0.14,  {2*\ystep}) {};
    \node[state] (T4b1) at (0.70,  {2*\ystep}) {};
    \node[state] (T4b2) at (1.26,  {2*\ystep}) {};
    \node[state] (T4b3) at (1.82,  {2*\ystep}) {};
    \draw[succ] (T4a0) -- (T4b0); \draw[succ] (T4a1) -- (T4b0);
    \draw[succ] (T4a2) -- (T4b1); \draw[succ] (T4a3) -- (T4b1);
    \draw[succ] (T4a4) -- (T4b2); \draw[succ] (T4a5) -- (T4b2);
    \draw[succ] (T4a6) -- (T4b3); \draw[succ] (T4a7) -- (T4b3);
    \node[discarded] (T4d0) at (2.38, {2*\ystep}) {};
    \node[discarded] (T4d1) at (2.94, {2*\ystep}) {};
    \node[discarded] (T4d2) at (3.50, {2*\ystep}) {};
    \node[discarded] (T4d3) at (4.06, {2*\ystep}) {};
    \draw[fail] (T4a8)  -- (T4d0); \draw[fail] (T4a9)  -- (T4d0);
    \draw[fail] (T4a10) -- (T4d1); \draw[fail] (T4a11) -- (T4d1);
    \draw[fail] (T4a12) -- (T4d2); \draw[fail] (T4a13) -- (T4d2);
    \draw[fail] (T4a14) -- (T4d3); \draw[fail] (T4a15) -- (T4d3);
    \node[state] (T4c0) at (0.42, {3*\ystep}) {};
    \draw[succ] (T4b0) -- (T4c0); \draw[succ] (T4b1) -- (T4c0);
    \node[discarded] (T4dd) at (1.54, {3*\ystep}) {};
    \draw[fail] (T4b2) -- (T4dd); \draw[fail] (T4b3) -- (T4dd);
    \node[rus] (T4r) at (0.42, \rusy) {RUS$_4$};
    \draw[succ] (T4c0) -- (T4r);
    \node[trial] at (2.10, -0.9) {$i=4$};
\end{scope}

\begin{scope}[xshift=10.1cm]
    \foreach \i in {0,...,15} {
        \node[state] (T5a\i) at ({0.28*\i}, {2*\ystep}) {};
    }
    \node[state] (T5b0) at (0.14, {3*\ystep}) {};
    \node[state] (T5b1) at (0.70, {3*\ystep}) {};
    \node[state] (T5b2) at (1.26, {3*\ystep}) {};
    \node[state] (T5b3) at (1.82, {3*\ystep}) {};
    \draw[succ] (T5a0) -- (T5b0); \draw[succ] (T5a1) -- (T5b0);
    \draw[succ] (T5a2) -- (T5b1); \draw[succ] (T5a3) -- (T5b1);
    \draw[succ] (T5a4) -- (T5b2); \draw[succ] (T5a5) -- (T5b2);
    \draw[succ] (T5a6) -- (T5b3); \draw[succ] (T5a7) -- (T5b3);
    \node[discarded] (T5d0) at (2.38, {3*\ystep}) {};
    \node[discarded] (T5d1) at (2.94, {3*\ystep}) {};
    \node[discarded] (T5d2) at (3.50, {3*\ystep}) {};
    \node[discarded] (T5d3) at (4.06, {3*\ystep}) {};
    \draw[fail] (T5a8)  -- (T5d0); \draw[fail] (T5a9)  -- (T5d0);
    \draw[fail] (T5a10) -- (T5d1); \draw[fail] (T5a11) -- (T5d1);
    \draw[fail] (T5a12) -- (T5d2); \draw[fail] (T5a13) -- (T5d2);
    \draw[fail] (T5a14) -- (T5d3); \draw[fail] (T5a15) -- (T5d3);
    \node[state] (T5c0) at (0.42, {4*\ystep}) {};
    \draw[succ] (T5b0) -- (T5c0); \draw[succ] (T5b1) -- (T5c0);
    \node[discarded] (T5dd) at (1.54, {4*\ystep}) {};
    \draw[fail] (T5b2) -- (T5dd); \draw[fail] (T5b3) -- (T5dd);
    \node[rus] (T5r) at (0.42, \rusy) {RUS$_5$};
    \draw[succ] (T5c0) -- (T5r);
    \node[trial] at (2.10, -0.9) {$i=5$};
\end{scope}

\draw[qubit] (T1r.east) -- node[above, font=\tiny]{fail} (T2r.west);
\draw[qubit] (T2r.east) -- node[above, font=\tiny]{fail} (T3r.west);
\draw[qubit] (T3r.east) -- node[above, font=\tiny]{fail} (T4r.west);
\draw[qubit] (T4r.east) -- node[above, font=\tiny]{fail} (T5r.west);
\draw[qubit] (T5r.east) -- ++(0.55, 0)
    node[right, font=\tiny]{$\cdots$};

\begin{scope}[xshift=-1.5cm, yshift=-1.7cm]
    \node[state] at (0.0, 0) {};
    \node[font=\small, anchor=west] at (0.22, 0) {magic state};
    \node[discarded] at (2.70, 0) {};
    \node[font=\small, anchor=west] at (2.92, 0) {discarded $\ket{+}$};
    \draw[succ] (5.55, 0) -- (6.05, 0);
    \node[font=\small, anchor=west] at (6.12, 0) {successful fusion};
    \draw[fail] (9.20, 0) -- (9.70, 0);
    \node[font=\small, anchor=west] at (9.77, 0) {failed fusion};
    \node[rus, minimum width=6.5mm, minimum height=4mm] at (12.55, 0)
        {\scriptsize RUS};
    \node[font=\small, anchor=west] at (13.00, 0)
        {teleport onto data qubit};
\end{scope}

\end{tikzpicture}
    \caption{Schematic of the balanced fusion RUS process for implementing a single-qubit rotation $R_z(\phi)$.  At trial $i$, we attempt to teleport a resource state $\ket{M(2^{i-1}\phi)}$ onto the data qubit. This resource state is built by fusing (in expectation) $4^{\lceil (i-1)/2\rceil}$ copies of $\ket{M(2^{\lfloor (i-1)/2\rfloor}\phi)}$ in a binary tree. Each fusion succeeds with probability $1/2$ (solid green edges) and otherwise yields $\ket{+}$, which is discarded (dashed red edges). On average, four inputs are required to produce one successful fusion output. If the RUS teleportation step (yellow box) at trial $i$ fails, we proceed to trial $i+1$, where the target angle has doubled.\label{fig:balanced_fusion}}
\end{figure*}
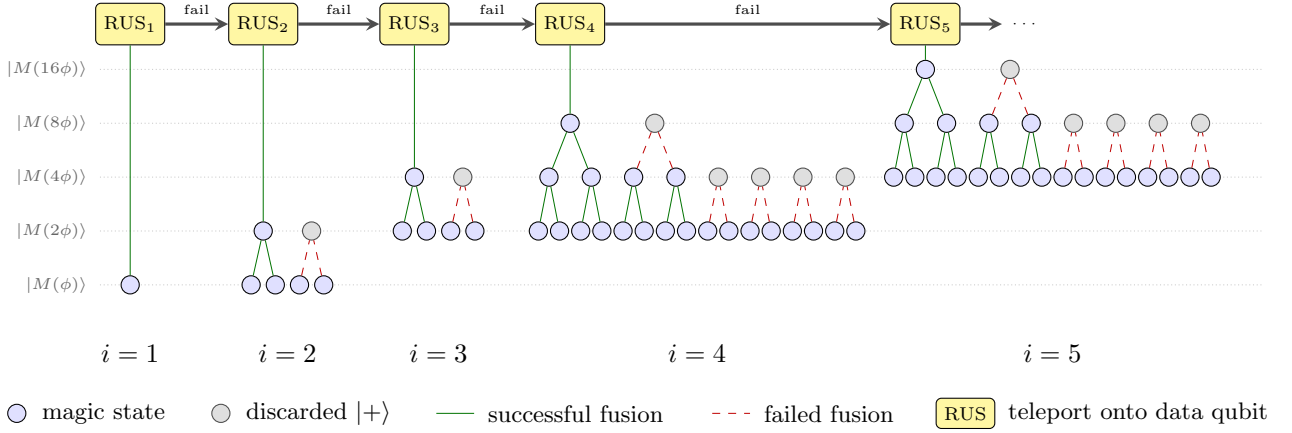

We assume access to a black-box protocol which prepares $\ket{M(\phi)}$ for arbitrary $\phi$. As discussed above, existing analog logical rotations proposals~\cite{choi2023rotations,toshio2025PartiallyFT} can prepare states that approximate $\ket{M(\phi)}$ up to  infidelity $\mathcal{O}(p \phi^2)$ and trace distance $\mathcal{O}(p \phi)$. In App.~\ref{App:MitigateCoherentErrors} we show that when each step of RUS is implemented as a mixture of rotation angles, the overall error is governed by the second moment of the angle error (i.e., the expectation of the squared deviation from the target angle)---which agrees with the infidelity rather than the trace distance. Moreover, even though the infidelity normally does not obey a triangle inequality (due to coherent error), in the context of RUS, the error accrues additively at each step, as if it obeyed a triangle inequality. Essentially, the teleportation steps in the RUS protocol ensure that the coherent error is decohered. As such, we will refer to the magic states as `having error' $\mathcal{O}(p \phi^2)$, although in reality it is more accurate to say that this is how the error accumulates, because of the decoherence of coherent error terms.

As discussed above, the baseline RUS process directly prepares magic states of the form $\ket{M(2^{m}\phi)}$, with error $\mathcal{O}(2^{2m}\phi^2p)$. We could instead consider preparing a number of magic states $\ket{M(\phi)}$, and `fuse' pairs via teleportation of one onto the other. If the teleportation succeeds, with probability $1/2$, we prepare $\ket{M(2\phi)}$. If the teleportation fails we obtain $\ket{+}$, and we discard the state. On average it takes 2 trials (thus 4 $\ket{M(\phi)}$ states) to fuse a pair. Once we have prepared states $\ket{M(2\phi)}$, we fuse these in pairs to produce copies of $\ket{M(4\phi)}$, and so on. To prepare the state $\ket{M(2^m\phi)}$ we require on average $4^m $ states. While this approach controls the error accumulation, it leads to a high resource cost.

We can improve on the protocol above by attempting to balance the error accumulation and resource state cost. Rather than building up a state $\ket{M(2^m\phi)}$ from $4^m$ states $\ket{M(\phi)}$ (on average), we could use states $\ket{M(2^{\lfloor \frac{m}{2}\rfloor }\phi)}$ as the basic fusion inputs. We require $4^{\lceil \frac{m}{2}\rceil }$ such resource states. See Fig.~\ref{fig:balanced_fusion} for a pictorial representation of this idea.  The expected error in the final resource state scales as $(2^{\lfloor \frac{m}{2}\rfloor }\phi)^2 p \times 2^{\lceil \frac{m}{2}\rceil } = 2^{\lfloor \frac{3m}{2}\rfloor }\phi^2 p$. If every resource state has unit cost, then the expected cost after $k$ trials of the RUS protocol with balanced fusion is:
\begin{equation}\label{eq:cost_balanced_fusion_k_trials}
    \sum_{i=1}^k \underbrace{\frac{1}{2^{i-1}}}_{\substack{\text{prob of} \\ \text{reaching trial $i$}}} \underbrace{4^{\lceil \frac{i-1}{2} \rceil }}_{\substack{\text{cost of}\\\text{trial $i$}}}  \leq   1.5 k.
\end{equation}
We note that our choice to fuse states from $\ket{M(2^{\lfloor \frac{m}{2}\rfloor }\phi)}$ is because this is the critical point that maintains sub-exponential scaling of resource costs (in $k$), while minimizing the average error. The expected accumulated error of the channel over $k$ trials is:
\begin{equation}\label{eq:error_balanced_fusion_k_trials}
    \sum_{i=1}^k \underbrace{\frac{1}{2^{i-1}}}_{\substack{\text{prob of} \\ \text{reaching trial $i$}}} \underbrace{  2^{\lfloor \frac{3(i-1)}{2} \rfloor} \phi^2 p  }_{\substack{\text{error in}\\\text{trial $i$}}}  < 3 p\phi^2 2^{0.5(k-1)}  
\end{equation}
One could consider tuning the depth of the fusion tree for tradeoffs between resource cost and error accumulation. 

Consider the special case where $\phi = \frac{\pi}{2^{k+2}}$. After each RUS trial, the angle is doubled, such that if $k$ trials fail, the target angle is $\pi/4$, which can be implemented using a Clifford $S$ gate. Because $k = \log_2\left(\frac{\pi}{4\phi} \right)$, the expected accumulated error in the channel is upper bounded by $\frac{3\sqrt{2\pi}}{4} \phi^{1.5}p$. In other words, the error scales \textit{better than linearly} with the target rotation angle. 

\begin{figure*}
    \centering
    \includegraphics[width=0.85\linewidth]{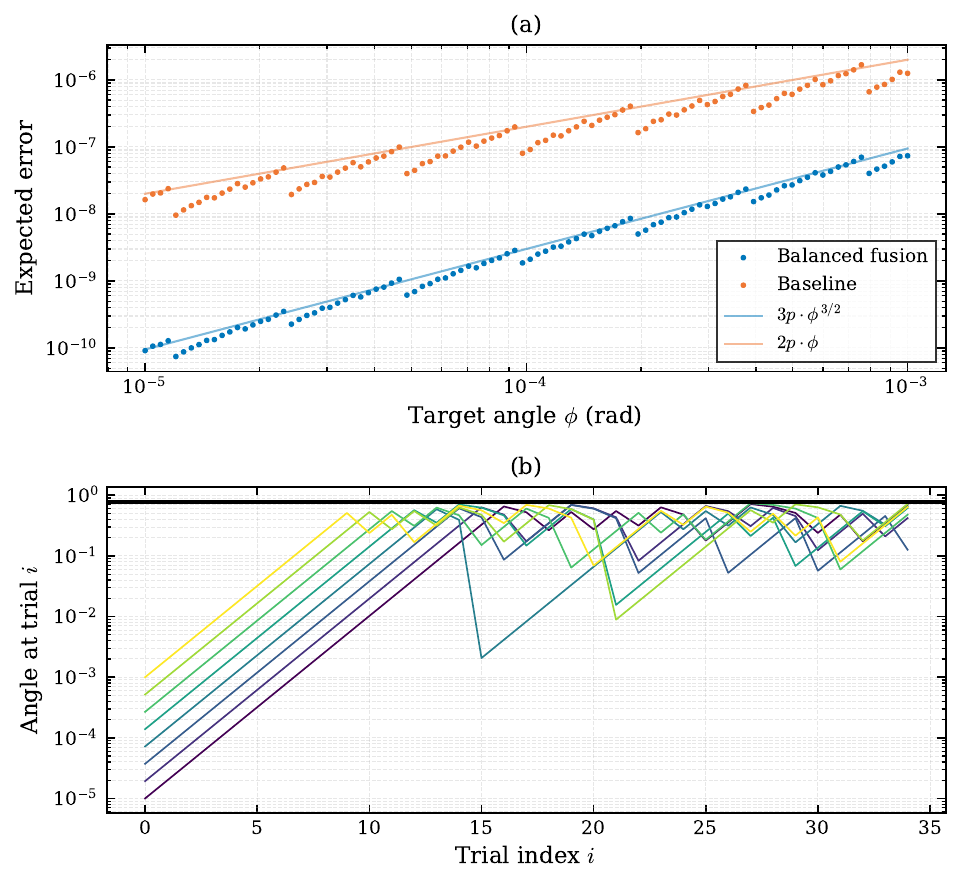}
    \caption{Error accumulation and angle growth during the RUS process with $i=35$ RUS trials and $p=10^{-3}$. Note that the probability of requiring more trials than this is $\leq 10^{-10}$.
    a) The expected error accumulated in the RUS process. When target angles exceed $\pi/4$, they are rescaled using an $S$ gate. Balanced fusion controls the error accumulation more effectively than the baseline approach. b) The growth of the target angle over the RUS iterations, for 100 different initial angles. Initially the target angle doubles on each iteration. When it would exceed $\pi/4$ (black horizontal line) it is rescaled. We observe that the target angle oscillates in size, but is typically much larger than the original target angle.}
    \label{fig:RUS_acc}
\end{figure*}

In general, $\phi \neq \pi/2^k$ for integer $k$. In that case, repeatedly doubling the angle will eventually cause it to become greater or equal to $\pi/4$ beginning on trial $i^* = \lceil \log_2(\pi/2\phi) \rceil$. When this happens, we can subtract $\pi/4$ using an $S$ gate, and then continue to implement the new target angle using balanced fusion RUS---in general, if on trial $i \geq i^*$ we have target angle $\phi' < \pi/4$, balanced fusion prepares $\ket{M(\phi')}$ by fusing an expected $4^m$ base states of angle $\phi'2^{-m}$, where $m =\lceil \frac{1}{2}\log_2(\phi'/\phi)\rceil $. This choice ensures that the error on the fused resource state is upper bounded by $2^{-m}\phi'^2p \leq \phi'^{1.5}\phi^{0.5}p < (\pi/4)^{1.5}\phi^{0.5}p$ and the expected cost by $4^m \leq \frac{4\phi'}{\phi} < \frac{\pi}{\phi}$. We show an example of this process in Fig.~\ref{fig:RUS_acc}. Notably, even though the target angles of each RUS step become much larger than the initial rotation angle, balanced fusion is able to control the accumulation of error. We observe that the total expected error is proportional to $\phi^{1.5}$, similar to that achieved for the special case above, and asymptotically improving over the baseline approach. The numerical observation of Fig.~\ref{fig:RUS_acc} may also be seen analytically by extending the sum in Eq.~\eqref{eq:error_balanced_fusion_k_trials} to $k=\infty$ but replacing the error for trial $i$ by $\min(2^{\lfloor \frac{3(i-1)}{2}\rfloor} \phi^2 p, (\pi/4)^{1.5}\phi^{0.5} p)$. There are two contributions: trials from $i=1$ to $i=i^*-1$ cumulatively contribute at most $(3\sqrt{\pi}/2)p\phi^{1.5}$, and trials from $i = i^*$ to $\infty$ cumulatively contributing at most $2^{-i^*+2}(\pi/4)^{1.5}p\phi^{0.5}\leq \sqrt{\pi}p\phi^{1.5}$, meaning the total error is upper bounded by $(5\sqrt{\pi}/2)p \phi^{1.5}$. Similarly, we may compute the expected cost by extending Eq.~\eqref{eq:cost_balanced_fusion_k_trials} to $k = \infty$ and replacing the cost of trial $i$ by $\min(4^{\lceil \frac{i-1}{2}\rceil},\frac{\pi}{\phi})$. The first $i^*-1$ trials contribute a cost $1.5(i^*-1)\leq 1.5 \log_2(\pi/2\phi)$ and the remaining trials contribute a cost $2^{-i^*+1}\frac{\pi}{\phi} \leq 4$, for a total cumulative cost of order $\log_2(1/\phi)$.

This calculation and its conclusion---that the leading order error of balanced fusion RUS is $\mathcal{O}(p \phi^{1.5})$---is derived more carefully in App.~\ref{App:MitigateCoherentErrors}. There, it is also observed that balanced fusion RUS has a second-order contribution of size $\mathcal{O}(p^2 \phi)$, deriving from the accumulation of coherent errors during the fusion process, prior to their natural decoherence in RUS. This first order term will dominate the second except when $\phi \lesssim p^2$. Moreover, the second term can be mitigated by engineering cancellation of coherent errors via a probabilistic rotation scheme, similar to \cite{toshio2025PartiallyFT}, discussed further in App.~\ref{App:ProbRots}.

\section{Applications}\label{Sec:Apps}

\subsection{Idealized unbounded Trotter evolution}

Consider a first-order Trotter evolution for time $\mathcal{T}$, using $r$ Trotter steps. We wish to simulate the longest time duration possible, subject to a limit on the algorithmic and logical errors. The error from Trotter scales as $\epsilon_{ts} \propto \mathcal{T}^2/r$~\cite{childs2021theory}. Consider a simplifying assumption of negligible memory or logical Clifford errors, such that the only logical errors arise from logical rotation gates. Using the baseline RUS approach, the logical error accumulation is $\epsilon_b \propto r \phi \propto \mathcal{T} $. Therefore the simulation is limited by error accumulation from logical rotation gates, and we can simulate a maximum time of $\mathcal{O}(1)$.

If instead logical rotation gates are implemented using balanced fusion RUS, then the logical error accumulation is $\epsilon_f \propto r \phi^{1.5} \propto \mathcal{T}^{1.5} / r^{0.5}$. If we choose $r=\Omega(\mathcal{T}^3)$, such that $\phi \propto 1/\mathcal{T}^2$, then we can arbitrarily suppress both Trotter and logical errors. This means that the maximum simulation time can be made unbounded by choosing an ever larger number of Trotter steps.

\subsection{More realistic analysis}
The arguments above are overly simplistic, for a number of reasons. First, memory and Clifford errors cannot be ignored, and will set a logical error floor that limits the maximum value of $r$, and hence the maximum evolution time $\mathcal{T}$. Second, our analysis of balanced fusion RUS relies on being able to prepare resource states with error $\mathcal{O}(p\phi^2)$ for any value of $\phi$. This assumption holds for the idealized analysis in App.~\ref{App:Overview}. However, if the noise model varies from simple depolarizing and includes realistic calibration limits on physical $R_z$ gates, then there may be an error floor on our resource states.

Nevertheless, we view analog logical rotations with balanced fusion RUS as a method for extending the maximum achievable simulation time on a processor with fixed spatial footprint, while maintaining polynomial time overhead. As we will show below, balanced fusion RUS may be capable of achieving logical error rates (and thus circuit depths) potentially inaccessible to cultivation + synthesis, or the baseline RUS approach. We will consider a protocol with fallback, where balanced fusion RUS is applied up to a number of trials $\kappa$, and then cultivation + synthesis is used to implement the corrective rotation if required. Magic state cultivation is able to produce $T$ states with an error of $10^{-9}$, using $10^5$ qubitrounds~\cite{gidney2024magic}. Suppressing the error with cultivation or distillation incurs an exponential overhead when they are operated using postselection, as is typically considered. Restricting to cultivation, we note that $R_z$ synthesis typically requires $50$--$100$ $T$ gates without the use of an ancilla qubit~\cite{ross2016optimal}. If an ancilla qubit is available, the cost of synthesis can be lowered by a factor of approximately $2$--$5$~\cite{Kliuchnikov2023shorterquantum}. As a result, cultivation + synthesis may struggle to implement $R_z(\phi)$ gates with errors below $10^{-8}$. In contrast, analog logical rotations with balanced fusion RUS and fallback (to cultivation + synthesis) may be able to reach lower target error rates. For example, if $\phi = 10^{-4}, p=10^{-3}, \kappa=8$, and setting synthesis error $\epsilon_s\approx 4\times 10^{-9}$ and $T$ state error $\epsilon_T \approx 10^{-9}$, then, working from Eq.~\eqref{eq:error_balanced_fusion_k_trials}, the expected error per $R_z(\phi)$ gate is:
\begin{align}
    &3 \alpha p\phi^2 2^{0.5(\kappa-1)} + 2^{-\kappa} \left(3 \log_2(1/\epsilon_s) \epsilon_T + \epsilon_s \right) \\
    &\approx (3.4 \times 10^{-10})\alpha + (3.4 \times 10^{-10}). 
\end{align}
The value of the unspecified constant $\alpha$ depends on the constant factor of the error for preparing each resource state $\ket{M(\phi)}$ (which depends on the quantum error correcting code and physical circuit used), as well as any additional constant factors of the error incurred during balanced fusion and RUS teleportation. These latter factors should be negligible if the code distance is sufficiently large that logical errors during Clifford gates and memory are subleading. In Ref.~\cite{toshio2025PartiallyFT}, the constant factor of the error for preparing $\ket{M(\phi)}$ was estimated as $d/30$ for a distance $d$ surface code under circuit level noise. Hence, it is reasonable to assume $\alpha \approx 1$, and we see that our method may reach an error regime approximately $15\times$ lower than cultivation + synthesis alone.

We can also evaluate the savings of balanced fusion + fallback, relative to the baseline RUS with fallback recently independently proposed in Ref.~\cite{toshio2026starmutation}. Relative to baseline RUS with fallback, balanced fusion achieves an error reduction of $2^{0.5(\kappa+1)}/3$ (see Eq.~(\ref{Eq:RUS_acc}) and Eq.~(\ref{eq:error_balanced_fusion_k_trials})), and requires a resource overhead of $0.75\kappa$ (based on Eq.~(\ref{eq:cost_balanced_fusion_k_trials}), and that 2 trials are required, on average, for baseline RUS). Note that this resource overhead assumes a unit cost for all resource states. In reality, for the proposal in App.~\ref{App:Overview}, the resource state cost increases substantially as $\phi$ increases, because of the reduction in postselection probability as $\phi$ increases. This further favours balanced fusion, which uses smaller-angle resource states than the baseline method.
\begin{itemize}
    \item For a realistic target regime with $\kappa = 6$, $\phi = 10^{-4}$, balanced fusion reduces the error by $3.8\times$ while incurring a resource overhead of $4.5\times$.
    \item For an optimistic target regime with $\kappa = 12$, $\phi = 10^{-5}$, balanced fusion reduces the error by $30\times$ while incurring a resource overhead of $\leq 9\times$.
\end{itemize}

The optimal value of $\kappa$ will depend sensitively on the protocols used for preparing $T$-states, and analog logical rotation magic states, which in turn depend on the assumptions made about the underlying architecture. We leave for future work a more complete resource analysis to determine which candidate early fault-tolerant Trotter simulations may benefit most from the improved error suppression of balanced fusion.

\section{Conclusion}
In this work we introduced the balanced fusion RUS method for reducing the error rate of analog logical rotations. While each logical analog rotation can be probabilistically implemented with an error of $\mathcal{O}(p \phi^2)$, the resulting error accumulation in RUS gate teleportation leads to worse overall error scaling. By balancing error accumulation and resource state overhead during the RUS gate teleportation process, our technique is able to asymptotically improve the $\mathcal{O}(p\phi)$ scaling of prior work to $\mathcal{O}(p\phi^{1.5})$. This improved error scaling may increase the accessible circuit depth in early fault-tolerant devices. 

There are a number of open questions regarding balanced fusion RUS, and the analog logical rotations technique in general:
\begin{itemize}[noitemsep]
    \item An immediate follow-up direction would be to perform more detailed resource analysis of balanced fusion RUS, comparing it to cultivation + synthesis for a realistic device error model. 
    \item Our current analysis provides bounds on the accumulation of errors in balanced fusion RUS, based on an idealized error model for the analog logical rotation magic states. An important follow up would be to implement a full noisy simulation of analog logical rotation magic state preparation, combined with balanced fusion and RUS teleportation. This will establish whether the assumptions in our analysis hold in practice.
    \item While there have been some analyses of non-depolarizing error models~\cite{toshio2025PartiallyFT,zeng2025errorstructuretailoredearlyfaulttolerantquantum,chung2026partiallyMegaQuop}, it would be valuable to investigate more realistic models for implementing logical analog rotations, such as considering finite physical $R_z$ calibration precision. The aim should be to identify the presence of logical error floors that could limit the applicability of the logical analog rotations technique. For example, consider the case where the error model is over-rotation by $R_z(x) \rightarrow R_z(\lambda x)$, for $\lambda=1+\xi$ in the physical rotation gates used to prepare $ \ket{M(\phi)}$ (see App.~\ref{App:Overview} for more details). This results in a magic state $\ket{M(\tilde{\phi})}$ with $\tilde{\phi} \approx \lambda^d x^d$ in a distance $d$ code (using the scheme of Ref.~\cite{choi2023rotations}). Then the infidelity is $\mathcal{O}(d^2 \xi^2 \phi^2)$, and the trace distance is $\mathcal{O}(d \xi \phi)$. For a simple dephasing noise model, we would expect an infidelity of $\mathcal{O}(dp\phi^2)$. The factor of $d$ increase compared to a simple dephasing noise model would worsen the results outlined above.
    \item It would be interesting to consider benefits from biased noise error models. For example, erasure qubits~\cite{kubica2023erasure} suppress Pauli noise in favor of heralded erasure errors that would be compatible with the postselection scheme in App.~\ref{App:Overview}. This has previously been exploited to reduce resource overheads for magic state distillation/cultivation~\cite{jacoby2025magic,vaknin2025efficient}. An erasure error model would be particularly compelling if it could be combined with the error analysis of Ref.~\cite{zeng2025errorstructuretailoredearlyfaulttolerantquantum} which claims to give a logical error rate scaling as $\mathcal{O}(p^2)$, rather than $\mathcal{O}(p)$.
\end{itemize}

Our work has made analog logical rotations more resilient to noise, and thus increased their viability as a competitor to cultivation/distillation combined with synthesis. Nevertheless, magic state cultivation is already able to achieve low logical error rates with substantially reduced spacetime overhead, compared to distillation. Moreover, a preprint released shortly before this paper has shown that synthesis costs may be reduced by orders of magnitude for small angle rotations, when targeting modest synthesis error~\cite{bothe2026more}. At low target synthesis error (the regime considered in our direct comparison above), synthesis costs match Ref.~\cite{Kliuchnikov2023shorterquantum}. As a result, it is still unresolved whether analog logical rotations will outperform cultivation + synthesis in practice, especially when moving from idealized analyses to realistic device models.

\section*{Acknowledgments}

We thank Catherine Leroux for helpful feedback on this manuscript. We thank Oskar Painter, James Hamilton, Nafea Bshara, Peter DeSantis, and Andy Jassy for their involvement and support of the research activities at the AWS Center for Quantum Computing.

\bibliography{Bib}

\newpage 
\onecolumngrid
\appendix

\section{Overview of analog logical rotations}\label{App:Overview}
We will discuss the approach for preparing analog logical rotation magic states introduced in Refs.~\cite{choi2023rotations,toshio2025PartiallyFT}. The method applies to distance $d$ CSS quantum error correcting (QEC) codes. The code is initially prepared in the logical $\ket{+}_L$ state. The physical operations $\exp(i xZ_j)$ are then applied to physical qubits in the support of the $Z_L$ logical operator of the code. The resulting operation is:
\begin{align}
    \bigotimes_{j=1}^d &\left(\cos(x)I_j + i\sin(x)Z_j \right) \\
    = & \left( \cos^d(x)I + i^d \sin^d(x) Z_L \right) + ...
\end{align}
where $+ ...$ denotes higher order terms that mix physical $I$ and $Z$ operations. When applied to $\ket{+}_L$, this operation prepares the state
\begin{equation}\label{Eq:StateBeforeProj}
    \sqrt{c^{2d} + s^{2d}} \ket{M(\phi)} + ...
\end{equation}
where $c = \cos(x), s=\sin(x)$, and
\begin{align}
    \ket{M(\phi)} &= \frac{c^d \ket{+}_L + i^d s^d \ket{-}_L}{\sqrt{c^{2d} + s^{2d}} } \\
    &:= \cos(\phi) \ket{+}_L + i^d \sin(\phi)\ket{-}_L \\
    &= \frac{1}{\sqrt{2}} \left( e^{i \phi} \ket{0}_L + e^{-i\phi}\ket{1}_L \right)
\end{align}
where the last equality holds when $d \equiv 1 \bmod 4$ (the equality can be enforced in other cases using additional Clifford gates). We have $\tan(\phi) = \tan^d(x)$. This implies $\phi \sim x^d$ for small $x$.

The additional terms in Eq.~(\ref{Eq:StateBeforeProj}) represent states outside of the codespace of the QEC code. As a result, they are flagged by nontrivial measurements of the $X$ stabilizers of the code. In the absence of physical errors, we obtain the trivial stabilizer outcome (flagging $\ket{M(\phi)}$) with probability
\begin{equation}
    P_s^0 = \cos^{2d}(x) + \sin^{2d}(x).
\end{equation}
Leaving the codespace means that the process is susceptible to logical errors caused by even a single physical $Z$ error, which can move an erroneous state back into the codespace. For example, the first-order erroneous state after such a $Z$ error is
\begin{equation*}
    isc \sqrt{c^{2d-4} + s^{2d-4}} \left( \cos(\phi_1) \ket{+}_L + i^{d-2} \sin(\phi_1) \ket{-}_L \right),
\end{equation*}
where $\tan(\phi_1) = \tan^{d-2}(x)$. If physical $Z$ errors occur with probability $p$, then the state $\ket{M(\phi_1)} = \cos(\phi_1) \ket{+}_L + i^{d-2} \sin(\phi_1) \ket{-}_L $ is prepared by the above process with probability $\mathcal{O}(p s^2 c^2 (c^{2d-4} + s^{2d-4}))$, and is the leading-order error process. In order to suppress measurement errors on the stabilizer measurements, we can repeat stabilizer measurements and postselect on repeatedly measuring the trivial stabilizers. Note that other sources of errors, such as physical $X$ errors, are corrected by the QEC code itself. The resulting density matrix after stabilizer measurement is a mixture of pure states. 

The output logical density matrix has the form 
\begin{equation}\label{Eq:MixtureOfRotations}
    q_0\ket{M(\phi)}\bra{M(\phi)} + \sum_{j \neq 0} q_j \ket{M(\phi_j)}\bra{M(\phi_j)}
\end{equation}
where $q_j$ is the probability of measuring a state $j$ phase flips from the codespace to have trivial stabilizer outcome. The density matrix can be re-expressed in the orthonormal basis $\ket{M(\phi)}, \ket{M(\phi)^\perp}$ where
\begin{equation}
    \ket{M(\phi)^\perp} = i\sin(\phi)\ket{+}_L + \cos(\phi)\ket{-}_L = Z_L \ket{M(\phi)}.
\end{equation}
Defining $\rho_\phi = \ket{M(\phi)}\bra{M(\phi)}$ and $\phi_0 = \phi$, the output density matrix is given by
\begin{equation}\label{Eq:OrthonormalDecomp}
    (\sum_{j} q_j \cos^2(\phi_j - \phi)) \rho_\phi + \sum_{j \neq 0} q_j \bigg(\sin^2(\phi_j - \phi)Z_L\rho_\phi Z_L
    + \frac{i}{2}\sin(2(\phi_j -\phi)) ( Z_L\rho_\phi - \rho_\phi Z_L) \bigg).
\end{equation}
In App.~\ref{App:InfidelityTraceDistance} we compute the infidelity between the output density matrix and $\rho_{\phi}$, without making the small $x$ approximation. For $\phi \ll 1, pd \ll 1,  \phi < pd < \phi^{2/d}$, the infidelity scales as $\mathcal{O}(p\phi^{2(1-\frac{1}{d})})$.

The resulting noisy magic state can be used to implement a logical rotation approximating $R_z(\phi)$. The magic state is consumed by gate teleportation, which succeeds with probability $1/2$. If the teleportation fails, a corrective operation $R_z(2\phi)$ is required. This naturally leads to a repeat-until-success (RUS) procedure for implementing $R_z(\phi)$ using the teleportation of analog logical rotation magic states.\\

Ref.~\cite{toshio2025PartiallyFT} modified the scheme outlined above to increase $P_s^0$, and to mitigate the reduction in fidelity from physical $Z$ errors. The idea is to group physical qubits in the support of $Z_L$, into $k$ groups of size $m$ (above, $k=d$ and $m=1$). Then rather than applying $R_z(x)^{\otimes d}$, we apply $R_{z^{\otimes m}} (x)^{\otimes k} $. For example, if $d=2k$ (with $k$ odd) and $m=2$, then we apply $R_{zz}(x)^{\otimes k}$. In all formulae above we can replace $d$ with $k$. The first benefit is that this increases the success probability $P_s^0$. The second benefit is that now a single physical $Z_j$ error cannot lead to a logical error. This is because the lowest order state outside of the codespace generated by $R_{zz}(x)^{\otimes k}$ is two phase-flips from the codespace. However, as noted in Ref.~\cite{toshio2025PartiallyFT}, the leading order error term is still $\mathcal{O}(p)$ under a depolarizing noise model, due to $Z_iZ_j$ error terms on the $R_{zz}(x)$ gates. Under alternative error models, the leading order error term scales as $\mathcal{O}(p^2)$~\cite{zeng2025errorstructuretailoredearlyfaulttolerantquantum}, but we will not consider this enhancement here. It was shown in Ref.~\cite{toshio2025PartiallyFT} that the infidelity of the output magic state scales as $\mathcal{O}\left(p \phi^{2(1-\frac{1}{k})}\right)$, while the trace distance scales as $\mathcal{O}(p \phi)$. Note, we approximate $2(1-1/k) \approx 2$ throughout this work. This trace distance scaling determines the error of the $R_z(\phi)$ gate implemented via gate teleportation, and so becomes the limiting factor.

\subsection{Calculation of the infidelity in the small logical angle limit}\label{App:InfidelityTraceDistance}
In this appendix, we will compute the infidelity between the output density matrix of the analog rotations protocol of Ref.~\cite{choi2023rotations} and the target magic state $\ket{M(\phi)}$, under an idealized noise model. The calculation is straightforward if one makes the small physical rotations angles approximation ($\sin(x) \sim x$). However, this approximation may not be justified for fixed, small $\phi$, because $x \rightarrow \pi/4$ as the code distance $d$ increases. We show that the same bound is obtained for $pd \ll 1$, $\phi \ll 1$, but the calculation is more laborious. 

We will work in the logical $\ket{+}, \ket{-}$ basis. In this basis, the state $\ket{M(\phi)}$ is given by
\begin{align}
    \ket{M(\phi)} = \frac{1}{\sqrt{\cos^{2d}(x) + \sin^{2d}(x)}}
    \begin{pmatrix}
        \cos^d(x) \\
        \left(i\sin(x)\right)^d
    \end{pmatrix}.
\end{align}

As in the previous subsection, we consider a distance $d$ CSS code, with $d$ odd. Begin in the $\ket{+}_L$ state. Apply $\exp(ix Z_j)$ to each physical qubit in the support of the $Z_L$ logical operator of the code. We then pass the resulting density matrix through a tensor product of simple dephasing channels $\mathcal{E}(\sigma) = (1-p)\sigma + p Z\sigma Z$, where $\sigma$ is a single-qubit density operator. We will only consider noise acting on the qubits in support of $Z_L$ here, as we will assume that the stabilizers are measured without introducing additional errors, and that physical errors on other qubits are corrected by the QEC code itself. Let $v \in \{0,1\}^d$ denote an error pattern where physical $Z$ errors occur. Then the state we obtain after measuring the trivial stabilizer is
\begin{align}
     \left(\cos^{d-|v|}(\theta)\sin^{|v|}(\theta)i^{|v|}\ket{+}_L + i^{d-|v|}\cos^{|v|}(\theta)\sin^{d-|v|}(\theta)\ket{-}_L\right).
\end{align}
For an error pattern of weight $|v|=j$, the contribution to the unnormalised density matrix has the form
\begin{align}
    \begin{pmatrix}
        \cos^{2d-2j}(x)\sin^{2j}(x)
        &
        (-i)^{d-2j}\cos^d(x)\sin^d(x)
        \\
        i^{d-2j}\cos^d(x)\sin^d(x)
        &
        \cos^{2j}(x)\sin^{2d-2j}(x)
    \end{pmatrix}.
\end{align}
Averaging over error weights gives the unnormalised state
\begin{align}
    \sum_{j=0}^d
    \binom{d}{j}p^j(1-p)^{d-j}
    \begin{pmatrix}
        \cos^{2d-2j}(x)\sin^{2j}(x)
        &
        (-i)^{d-2j}\cos^d(x)\sin^d(x)
        \\
        i^{d-2j}\cos^d(x)\sin^d(x)
        &
        \cos^{2j}(x)\sin^{2d-2j}(x)
    \end{pmatrix}.
\end{align}
Using the binomial theorem to compute the sum for each entry, the unnormalized density matrix is given by
\begin{align}
    \begin{pmatrix}
        \left(p+q\cos^2(x)\right)^d
        &
        \left(-iq\cos(x)\sin(x)\right)^d
        \\
        \left(iq\cos(x)\sin(x)\right)^d
        &
        \left(p+q\sin^2(x)\right)^d
    \end{pmatrix},
\end{align}
where $q = (1-2p)$.

By direct calculation, the fidelity with $\ket{M(\phi)}$ is
\begin{align}
   F =  \frac{
        \cos^{2d}(x)\left(p+q\cos^2(x)\right)^d
        +
        \sin^{2d}(x)\left(p+q\sin^2(x)\right)^d
        +
        2q^d\cos^{2d}(x)\sin^{2d}(x)
    }{
        \left[
            \left(p+q\cos^2(x)\right)^d
            +
            \left(p+q\sin^2(x)\right)^d
        \right]
        \left(\cos^{2d}(x)+\sin^{2d}(x)\right)
    }.
\end{align}
Recall the relation between the target logical rotation angle and the physical rotation angle
\begin{align}
    \tan(\phi)=\tan^d(x).
\end{align}
Let
\begin{align}
    t = \tan(\phi),
    \qquad
    a = t^{2/d}.
\end{align}
Then
\begin{align}
    \cos^2(x)=\frac{1}{1+a},
    \qquad
    \sin^2(x)=\frac{a}{1+a},
    \qquad
    \tan^2(\phi)=a^d.
\end{align}
Using
\begin{align}
    p+q\cos^2(x)
    ={}&
    p+\frac{q}{1+a}
    =
    \frac{1-p+pa}{1+a},
    \\
    p+q\sin^2(x)
    ={}&
    p+\frac{qa}{1+a}
    =
    \frac{p+(1-p)a}{1+a},
\end{align}
the common factor of $(1+a)^{-d}$ cancels between numerator and denominator. The fidelity becomes
\begin{align}
    F
    =
    \frac{
        (1-p+pa)^d
        +
        t^2\left(p+(1-p)a\right)^d
        +
        2q^d t^2
    }{
        \left[
            (1-p+pa)^d
            +
            \left(p+(1-p)a\right)^d
        \right]
        (1+t^2)
    }.
\end{align}
Hence the infidelity is
\begin{align}
    1-F
    ={}&
    \frac{
        \left(p+(1-p)a\right)^d
        +
        t^2(1-p+pa)^d
        -
        2q^d t^2
    }{
        \left[
            (1-p+pa)^d
            +
            \left(p+(1-p)a\right)^d
        \right]
        (1+t^2)
    }.
\end{align}
Expanding in the regime
\begin{align}
    pd \ll 1,
    \qquad
    \phi \ll 1,
    \qquad
    \phi < pd < \phi^{2/d},
\end{align}
we use 
\begin{align}
    (1-p+pa)^d
    ={}&
    (1-p(1-a))^d
    =
    1-dp(1-a)+O(p^2d^2),
    \\
    \left(p+(1-p)a\right)^d
    ={}&
    \left(a+p(1-a)\right)^d
    =
    a^d\left(1+dp\frac{1-a}{a}\right)
    +O(p^2d^2),
    \\
    q^d
    ={}&
    (1-2p)^d
    =
    1-2dp+O(p^2d^2).
\end{align}
Since $a^d=t^2$, this gives
\begin{align}
    1-F
    =
    dp\,
    \frac{
        t^2(1+a)^2
    }{
        a(1+t^2)^2
    }
    +
    O(p^2d^2).
\end{align}
Equivalently,
\begin{align}
    1-F
    =
    dp\,
    \frac{
        \tan^{2-2/d}(\phi)
        \left(1+\tan^{2/d}(\phi)\right)^2
    }{
        \left(1+\tan^2(\phi)\right)^2
    }
    +
    O(p^2d^2).
\end{align}
For $\phi\ll 1$, this reduces to
\begin{align}
    1-F
    =
    dp\,\phi^{2\left(1-\frac{1}{d}\right)}
    +
    O(p^2d^2).
\end{align}

\section{Natural mitigation of coherent errors}\label{App:MitigateCoherentErrors}

Here we rigorously justify that when each step of RUS is implemented as a mixture of rotation angles, the overall error is governed by the second moment of the angle error at each step, even when the first moment of angle error is nonzero (i.e., when there is coherent error). The second moment scales as the infidelity, whereas the trace error can be much larger due to the coherent error. Essentially, the teleportation steps in the RUS protocol ensure that the coherent error is decohered. For the balanced fusion approach specifically---as long as angles are reset once they exceed $\pi/4$, as described in the main text---the RUS scheme naturally achieves overall $\mathcal{O}(p\phi^{1.5})$ error scaling in trace distance, to first order in $p$, without any need for tailored correction cancellations.

\subsection{The Mixing Lemma}
As shown in Eq.~(\ref{Eq:MixtureOfRotations}), the protocols for preparing analog logical rotation magic states output a mixture of magic states for rotations by different angles. Teleporting in a state of this form (for use in RUS, or balanced fusion), will implement a channel representing a mixture of rotations. We can use the Mixing Lemma~\cite{hastings2016turning,campbell2017Mixing} to bound the error in this channel. The Mixing Lemma observes that for unitary channels $\mathcal{V}(\cdot) = V(\cdot)V^\dagger$ and $\{ \mathcal{U}_i(\cdot) = U_i (\cdot) U_i^\dagger \}$, where $\mathcal{U}_i$ are drawn from a probability distribution, 
\begin{equation}\label{Eq:Mixing}
    \frac{1}{2} \left\lVert \mathcal{V}(\cdot) - \E[ \mathcal{U}] (\cdot)\right\rVert_\diamond \leq 2 \left\lVert V - \E[ U] \right\rVert
\end{equation}
We refer the reader to Ref.~\cite{zhao2026exponential} for a detailed derivation. For the special case where $\{U_i \}$ are diagonal rotations by angles $\{\phi_i\}$, and $V$ is a diagonal rotation by the mean $\bar{\phi}$, we can show the following result:

Let $\bar{\phi} = \mathbb{E}(\phi_j)$ be the mean and $\delta = \phi_j - \bar{\phi}$ be a shifted random variable with mean zero. Then
\begin{align}\label{Eq:MixingVarBound}
    2 \lVert V - \mathbb{E}[U] \rVert &= 2\lvert e^{i \bar{\phi}} - \mathbb{E}[e^{i\phi_j}] \rvert \\ 
    &= 2\left\lvert 1- \mathbb{E}\left[e^{i\delta}\right] \right\rvert \\
    &= 2 \left\lvert 1-\mathbb{E}\left[1+i\delta + \int_{0}^\delta ds \int_{0}^s ds'(-e^{is'})\right]\right\rvert \qquad \text{Taylor expansion with remainder} \\
    &= 2\left\lvert i\underbrace{\mathbb{E}[\delta]}_{0} + \mathbb{E}\left[\int_{0}^\delta ds \int_{0}^s ds'(-e^{is'})\right] \right\rvert \\
    &\leq 2 \int_{0}^\delta ds \int_{0}^s ds'\mathbb{E}\left[\lvert e^{is'}\rvert\right]  \\
    &=  2\int_{0}^\delta ds \int_{0}^s ds'   \\
    &= \mathbb{E}[\delta^2] = \mathrm{Var}(\phi_j) 
\end{align}

In other words, if we approximate the mixture of rotations by a single rotation with the mean angle, then the approximation error is upper bounded by the variance of the rotation angles.

\subsection{RUS teleportation}

In RUS, ideally on trial $i$ we prepare the magic state $\ket{M(\phi^{(i)})}$, where $\phi^{(1)} = \phi$ and for $i \geq 2$
\begin{align}
    \phi^{(i)} = \begin{cases}
        2\phi^{(i-1)} & \text{if } \phi^{(i-1)} < \pi/8 \\ 2\phi^{(i-1)}-\pi/4 &
        \text{otherwise}
    \end{cases}
\end{align}
that is, we reset so that $\phi^{(i)} < \pi/4$ always holds. Let the rotation channel by angle $\theta$ be denoted $\mathcal{R}_z(\theta) := R_z(\theta)[\cdot] R_z(\theta)^\dagger$. Using perfect resource states, the channel implemented on trial $i$ of RUS with measurement outcome $b$ would be 
\begin{align}
\mathcal{R}^{(i)}_b = \mathcal{R}_z((-1)^{b} \phi^{(i)}) \circ \mathcal{S}^{(i)}
\end{align}
where $\mathcal{S}^{(1)} = \mathcal{I}$ and
\begin{align}
    \mathcal{S}^{(i)} = 
    \begin{cases}
        \mathcal{I} & \text{if } \phi^{(i-1)} < \pi/8 \\
        \mathcal{R}_z(\pi/4)& \text{otherwise}
    \end{cases}
\end{align}
is the $S$ (rotation by $\pi/4$) gate in the case that doubling the angle at step $i-1$ causes the angle to become greater or equal to $\pi/4$ and identity otherwise. 
The total RUS channel would then be given by
\begin{align}
    \mathcal{R}_{\rm RUS} = \sum_{i=1}^{\infty} \frac{1}{2^i} \mathcal{R}_0^{(i)}\circ \mathcal{R}_1^{(i-1)} \circ \mathcal{R}_1^{(i-2)} \circ \cdots \circ \mathcal{R}_1^{(1)}  = \sum_{i=1}^{\infty} \frac{1}{2^i}\mathcal{R}_z(\phi) = \mathcal{R}_z(\phi)
\end{align}
where the $2^{-i}$ factor represents the probability that we terminate on trial $i$. The equality with $\mathcal{R}_z(\phi)$ follows from the correctness of the RUS protocol and is seen by the fact that for each term in the sum, the total angle rotated is exactly $\phi$. 

In the schemes outlined in this paper, we approximate this by teleporting a mixed state of the form
\begin{align}
    \rho^{(i)} = q_0^{(i)} \ket{M(\phi^{(i)})}\bra{M(\phi^{(i)})} + \sum_{j \neq 0} q_j^{(i)} \ket{M(\phi^{(i)}_j)}\bra{M(\phi^{(i)}_j)} 
\end{align}
which is a mixture of magic states of different angles. Suppose we teleport this state and obtain measurement outcome $b$. Instead of implementing $\mathcal{R}_z((-1)^b\phi^{(i)})$, we instead implement $\mathcal{E}^{(i)}_b \circ \mathcal{R}_z((-1)^b\phi^{(i)})$, where the error channel is
\begin{align}
    \mathcal{E}^{(i)}_b &= 
     q_0^{(i)} \mathcal{I} + \sum_{j \neq 0} q_j^{(i)} \mathcal{R}_z((-1)^b(\phi^{(i)}_j-\phi^{(i)}))
\end{align}
which is a mixture of rotation channels. Note that mixtures of rotation channels commute with other mixtures of rotation channels, which encompasses all channels defined in this section. 
Thus, on trial $i$ of RUS with measurement outcome $b$, instead of implementing $\mathcal{R}^{(i)}_b$, we implement
\begin{align}
    \tilde{\mathcal{R}}_b^{(i)} =  \mathcal{E}_b^{(i)} \circ \mathcal{R}_z((-1)^{b} \phi^{(i)}) \circ \mathcal{S}^{(i)}
\end{align}
The total RUS channel is given by
\begin{align}
    \tilde{\mathcal{R}}_{\rm RUS} &= \sum_{i=1}^{\infty} \frac{1}{2^{i}} \tilde{\mathcal{R}}_{0}^{(i)} \circ \tilde{\mathcal{R}}_{1}^{(i-1)} \circ \tilde{\mathcal{R}}_{1}^{(i-2)} \circ \cdots \circ \tilde{\mathcal{R}}_{1}^{(1)} \\
    &= \sum_{i=1}^{\infty}\frac{1}{2^i} (\mathcal{E}_0^{(i)}\circ \mathcal{E}_1^{(i-1)} \circ \mathcal{E}_1^{(i-2)} \circ \cdots \circ \mathcal{E}_1^{(1)})\circ (\mathcal{R}_0^{(i)}\circ \mathcal{R}_1^{(i-1)} \circ \mathcal{R}_1^{(i-2)} \circ \cdots \circ \mathcal{R}_1^{(1)}) \\
    &= \underbrace{\left[\sum_{i=1}^{\infty}\frac{1}{2^i} (\mathcal{E}_0^{(i)}\circ \mathcal{E}_1^{(i-1)} \circ \mathcal{E}_1^{(i-2)} \circ \cdots \circ \mathcal{E}_1^{(1)})\right]}_{\mathcal{E}} \; \circ \; \mathcal{R}_z(\phi) \\
\end{align}
which is the correct channel followed by an error channel $\mathcal{E}$, which is a mixture of rotation channels. 
Define independent random variables $\Phi^{(i)}$ for $i=1,2,\ldots$ which are equal to the random angle rotated in rotation ensemble $\mathcal{E}_0^{(i)}$. Note  that $\mathbb{E}[\Phi^{(i)}] \neq 0$ in general---this represents the coherent error in each step of RUS. We define the random variable $\Phi$ by first choosing a random $i$ (with probability $2^{-i}$) and setting
\begin{align}
    \Phi = \Phi^{(i)} - \Phi^{(i-1)} - \Phi^{(i-2)} - \ldots - \Phi^{(1)}
\end{align}
so that $\Phi$ represents the random angle rotated in rotation ensemble $\mathcal{E}$. 
We can confirm that even though the mean of each $\Phi^{(i)}$ does not vanish, the mean of $\Phi$ does vanish:
\begin{align}
    \mathbb{E}[\Phi] &= \sum_{i=1}^\infty \frac{1}{2^i} \mathbb{E}[\Phi^{(i)} - \sum_{m=1}^{i-1} \Phi^{(m)}] \\
    &= \sum_{m=1}^{\infty} \mathbb{E}[\Phi^{(m)}](\frac{1}{2^m} - \sum_{i=m+1}^{\infty} \frac{1}{2^m}) = 0
\end{align}
We can also compute the variance
\begin{align}
    \mathrm{Var}[\Phi] &= \mathbb{E}[\Phi^2] = \sum_{i=1}^\infty \frac{1}{2^i} \mathbb{E}[(\Phi^{(i)} - \sum_{m=1}^{i-1} \Phi^{(m)})^2] \\
    &=\sum_{i=1}^{\infty}\frac{1}{2^i}\left(\sum_{m \leq i} \mathbb{E}[{\Phi^{(m)}}^2]-2\sum_{m < i} \mathbb{E}[\Phi^{(i)}]\mathbb{E}[\Phi^{(m)}] + 2\sum_{m' < m < i}\mathbb{E}[\Phi^{(m)}]\mathbb{E}[\Phi^{(m')}]\right) \\
    &= \sum_{m=1}^\infty \mathbb{E}[{\Phi^{(m)}}^2]\underbrace{\left(\sum_{i=m}^\infty \frac{1}{2^i}\right)}_{2^{-m+1}} + \sum_{m'=1}^{\infty} \sum_{m=m'+1}^{\infty}\mathbb{E}[\Phi^{(m)}]\mathbb{E}[\Phi^{(m')}] \underbrace{\left(\frac{-2}{2^m} +\sum_{i=m+1}^\infty \frac{2}{2^i}\right)}_{0} \\
    &= 2 \sum_{m=1}^{\infty} \frac{1}{2^m} \mathbb{E}[{\Phi^{(m)}}^2]
\end{align}
Using the mixing lemma result from above, we then have
\begin{align}\label{eq:full_error_RUS}
    \frac{1}{2}\lVert \tilde{R}_{\rm RUS} - R_{\rm RUS} \rVert_{\diamond} \leq  \mathrm{Var}[\Phi] \leq \sum_{m=1}^{\infty} 2^{-m+1} \mathbb{E}[{\Phi^{(m)}}^2]
\end{align}
Recall that for mixtures of rotation angles, the second moment of the angle scales as the infidelity, whereas the trace error can be much larger: given an ensemble of rotations described by a random angle $\Phi^{(i)}$, the infidelity relative to the identity channel is precisely
\begin{align}
    1-F = \mathbb{E}[\sin^2(\Phi^{(i)})] 
\end{align}
which falls within a constant factor of the second moment: 
\begin{align}
    \frac{4}{\pi^2}\mathbb{E}[{\Phi^{(i)}}^2] \leq 1-F \leq \mathbb{E}[{\Phi^{(i)}}^2]\,.
\end{align}
Above, the lower bound holds whenever we restrict $\Phi^{(i)} \in [-\pi/2, \pi/2]$ (where $\sin^2(\Phi^{(i)}) \geq (4/\pi^2){\Phi^{(i)}}^2$), which is always possible, as $\ket{M(\phi)} = \ket{M(\phi + \pi)}$ up to unphysical global phase.

\subsection{The second moment in balanced fusion}

For any angle $\theta$, the analog rotations method allows the preparation of base states $\rho = \mathcal{E}(\ket{M(\theta)}\bra{M(\theta)})$, where $\mathcal{E}$ is a noise channel which is close to the identity, and equal to a mixture of rotation angles with random angle $\Phi$ satisfying
\begin{align}
    \mathbb{E}[\Phi] &= \mathcal{O}(p\theta) \\
    \mathbb{E}[\Phi^2] &= \mathcal{O}(p\theta^2)
\end{align} 

When one successfully fuses two states $\rho$, one arrives at $\mathcal{E}(\mathcal{E}(\ket{M(2\theta)}\bra{M(2\theta)}))$. In general, after depth $t$ of successful fusions, one arrives at the state $\mathcal{E}^{2^t}(\ket{M(2^t\theta)}\bra{M(2^t\theta)})$. The total angle of rotation $\Phi_{\rm tot}$ of channel $\mathcal{E}^{2^t}$ is the sum of $2^t$ i.i.d.~variables with first and second moment as above; thus we have
\begin{align}
    \mathbb{E}[\Phi_{\rm tot}] &= 2^t \mathbb{E}[\Phi] = \mathcal{O}(p2^t\theta) \\
    \mathbb{E}[\Phi_{\rm tot}^2] &= 2^t \mathbb{E}[\Phi^2] + 2^t(2^t-1)\mathbb{E}[\Phi]^2 = \mathcal{O}(p2^{t}\theta^2 + p^2 2^{2t}\theta^2)
\end{align}

In balanced fusion, on trial $m < \lceil \log_2(\pi/2\phi)\rceil$, the target angle is $2^m \phi$. One uses base angle $\theta = 2^{\lfloor \frac{m}{2} \rfloor}\phi$ and performs depth $t = \lceil \frac{m}{2}\rceil$ of fusions. Thus, one has
\begin{align}
    \mathbb{E}[\Phi_{\rm tot}^2] = \mathcal{O}(p 2^{1.5m}\phi^2 + p^22^{2m}\phi^2) \qquad \qquad \text{when } m = 1,\ldots, \lceil \log_2(\pi/2\phi) \rceil -1
\end{align}
On trial $m \geq \lceil \log_2(\pi/2\phi)\rceil$, the target angle has been reset to some $\phi' < \pi/4$. One uses base angle $\theta = \phi'2^{-\lceil \frac{1}{2}\log_2(\phi'/\phi) \rceil } \leq \sqrt{\phi\phi'}$ and performs depth $t = \lceil \frac{1}{2}\log_2(\phi'/\phi) \rceil$ of fusions. Thus one has 
\begin{align}
    \mathbb{E}[\Phi_{\rm tot}^2] = \mathcal{O}(p \phi^{0.5}+ p^2) \qquad \qquad \text{when } m \geq \lceil \log_2(\pi/2\phi) \rceil
\end{align}

Applying the formula from \cref{eq:full_error_RUS}, we have that the total error is upper bounded by 
\begin{align}
 &\sum_{m=1}^{\lceil \log_2(\pi/2\phi) \rceil-1} 2^{-m+1}  \mathcal{O}(p 2^{1.5m}\phi^2 + p^22^{2m}\phi^2) + \sum_{\lceil \log_2(\pi/2\phi) \rceil}^\infty 2^{-m+1}  \mathcal{O}(p \phi^{0.5}+ p^2) \\
 \leq{}& \mathcal{O}(p\phi^{1.5} + p^2 \phi)
\end{align}

 For typical parameter ranges considered in this work (e.g. $p=10^{-3}$, $\phi > 10^{-6}$), the first term dominates. The second term arises from the residual coherent error in the magic states used in balanced fusion. This is present because $\E[\Phi^{(i)}] \neq 0$ in general. These coherent errors accumulate linearly during the fusion process, before being decohered by RUS teleportation. If the second term were problematic, it could be mitigated by using the analysis in App.~\ref{App:MixingLemma}. That analysis approximates the mixture of rotations (implemented by each magic state teleportation) by a single rotation with angle equal to the mean angle of the mixture. Using the Mixing Lemma, this can be shown to eliminate the problematic coherent error term.

\section{Comparison to prior work}\label{App:Comparison}

In this section, we compare our new techniques to prior work. In particular, we compare to randomization methods introduced in Ref.~\cite{toshio2025PartiallyFT}, and re-examine the twirling method introduced in Ref.~\cite{zhang2025low}.

\subsection{A short summary of recent progress}\label{App:Summary}

Analog logical rotations have been investigated in a number of recent publications, as we discuss below:

\begin{itemize}
    \item \cite{choi2023rotations}: This paper introduces the $m=1$ logical analog rotations method described in App.~\ref{App:Overview}, and shows that the infidelity of the magic state scales as $\mathcal{O}(p\phi^2)$.
    \item \cite{toshio2025PartiallyFT}: This work makes a number of contributions:
    \begin{enumerate}
        \item Introduces the $m>1$ logical analog rotations method described in App.~\ref{App:Overview}, which has a higher success probability and better error scaling than the $m=1$ case.
        \item Observes that while the infidelity of the magic state scales as $\mathcal{O}(p\phi^2)$, the trace distance scales as $\mathcal{O}(p\phi)$ due to coherent terms.
        \item Uses the probabilistic rotations (see App.~\ref{App:ProbRots}) to effectively cancel the coherent error. However, due to error accumulation in the RUS process (Eq.~(\ref{Eq:RUS_acc}) in our main text), the overall error in the gate is still $\mathcal{O}(p\phi)$.
    \end{enumerate}
    \item \cite{akahoshi2024compilation}: This paper applies the methods of Ref.~\cite{toshio2025PartiallyFT} to detailed resource estimates for QPE.
    \item \cite{ismail2025transversal}: This paper applies the analog logical rotations method to a higher-rate LDPC architecture suitable for realization on reconfigurable neutral atom arrays. 
    \item \cite{zhang2025low}: This paper introduces a twirling procedure, the analysis of which we revisit in detail in App.~\ref{App:Twirl}. The paper also proposes a method to concatenate the transversal rotation technique, first applying it on the physical level to produce encoded magic states, and then applying it at the logical level to further suppress the error in the output magic state.
    \item \cite{huang2025robust, yoshioka2025transversal}: These papers consider in-place implementation of the logical analog rotation. Rather than post-selecting on the trivial stabilizer outcome, these papers observe that any stabilizer outcome indicates that a known logical rotation has been implemented. If the initial logical rotation does not succeed, then the process can be repeated with an updated target angle. Because logical rotations with large $\phi_j$ occur with relatively high probability, the logical error rate scales as $\mathcal{O}(p \phi)$.
    \item \cite{zeng2025errorstructuretailoredearlyfaulttolerantquantum}: This paper considers the $m=2$ approach of Ref.~\cite{toshio2025PartiallyFT}, which uses physical gates of the form $\exp(i \theta ZZ)$. The paper gives justification that under certain superconducting qubit inspired noise models, the damaging two-qubit $ZZ$ errors only occur with probability $\mathcal{O}(p^2)$. This improves the error dependence in any of the $ m=2$ protocols~\cite{toshio2025PartiallyFT} from $\mathcal{O}(p)$ to $\mathcal{O}(p^2)$.
    \item \cite{chung2026partiallyMegaQuop}: This work considers the $m \geq 1$ proposals of Ref.~\cite{toshio2025PartiallyFT} under a realistic superconducting qubit noise model. Their numerical results indicate that in order to maintain a moderate success probability for preparing $\ket{M(\phi)}$, the code distance $d$ must not be too large. However, if $d$ is too small, logical memory errors in the surface code patch may result in an error floor that limits the performance of the analog logical rotations method. This work then analyzes the effect of a protocol to grow the surface code patches to the size used in the rest of the computation.
    \item \cite{toshio2026starmutation}: This recent work augments the RUS logical analog rotations method with fallback to Clifford + $T$ synthesis. If the number of RUS trials exceeds a specified threshold $k_{th}$, then the corrective operation is implemented via Clifford + $T$ synthesis. The benefit of the fallback protocol is that it eliminates the need to prepare large-angle corrective rotations with the analog logical rotations approach. Because fallback is rarely needed, the total resources required for synthesis in the circuit can be significantly reduced. 
    \item \cite{sethi2026injeqt}: This paper investigates the spacetime costs of using logical analog rotation magic states in an extractor-based Gross code architecture~\cite{yoder2025tour}. 
\end{itemize}

\subsection{Probabilistic rotations}\label{App:ProbRots}
In Ref.~\cite{toshio2025PartiallyFT}, a probabilistic rotation cancellation scheme was developed. It is observed that to leading order, the noise channel from successfully teleporting the state in Eq.~(\ref{Eq:MixtureOfRotations}) is
\begin{equation}
    \mathcal{E}_\phi(\rho) = (1-q_1) \rho + q_1\mathcal{R}_{\phi_1 - \phi}(\rho),
\end{equation}
where $\mathcal{R}_\theta$ is the channel representation of the rotation $R_z(\theta)$. Ref.~\cite{toshio2025PartiallyFT} then defines a corrective channel
\begin{equation}
    \mathcal{C}_\phi (\rho) = (1-q_1) \rho + q_1 \widetilde{\mathcal{R}}_{\phi - \phi_1}(\rho),
\end{equation}
where the tilde denotes that the probabilistic rotation is implemented by a noisy RUS teleportation process of its own. Note the change in sign of the rotation in the corrective channel, compared to the noise channel. The combined effect of the corrected noise channel is
\begin{equation}
    \mathcal{C}_\phi \circ \mathcal{E}_\phi(\rho) = (1-q_1)^2 \rho + 2(1-q_1)q_1 (\cos^2(\phi_1 - \phi) \rho + \sin^2(\phi_1 - \phi) Z\rho Z) + \mathcal{O}(\phi p^2).
\end{equation}
Note that the leading-order error term $\mathcal{O}(p\phi^2)$ may be comparable to the next-to-leading-order term $\mathcal{O}(\phi p^2)$. This issue was later addressed in Ref.~\cite{toshio2026starmutation}. The technique was extended to consider a corrective channel 
\begin{equation}
    \mathcal{C}'_\phi (\rho) = (1-\sum_{j\neq 0} q_j) \rho + \sum_{j \neq 0} q_j \mathcal{R}_{\phi - \phi_j}(\rho),
\end{equation}
which now 1) accounts for higher-order erroneous rotations, and 2) implements the probabilistic corrective rotations using cultivation + synthesis to eliminate accrual of coherent errors. This improves the next-to-leading-order error term to $\mathcal{O}(p^2 \phi^2)$. Note that these mitigation methods require additional overhead to implement the corrective rotations, and also require accurate calibration of $\{ q_j \}$.

\subsubsection{Improvements to the analysis of probabilistic rotations using the Mixing Lemma}\label{App:MixingLemma}

A more direct method could proceed as follows. Observe that teleporting in the state of Eq.~(\ref{Eq:MixtureOfRotations}) produces a mixture of unitary channels
\begin{equation}
    \sum_{j=0}^{\lfloor d/2 \rfloor} \tilde{q}_j \mathcal{U}_{\phi_j}
    +
    \tilde{q}'_j \mathcal{U}_{\phi'_j},
\end{equation}
where $d$ is odd, $\sum_j(\tilde{q}_j+\tilde{q}'_j)=1$, and $\mathcal{U}_{\phi}$ denotes conjugation by $\exp(i\phi Z)$. For completeness, we make the distribution explicit. Let
\begin{equation}
    m=\frac{d-1}{2}.
\end{equation}
The unnormalized probabilities are
\begin{align}
    q_j
    &=
    \binom{d}{j}
    \sin^{2j}(x)\cos^{2j}(x)
    \left(
        \sin^{2d-4j}(x)+\cos^{2d-4j}(x)
    \right)
    p^j(1-p)^{d-j},
    \\
    q'_j
    &=
    \binom{d}{j}
    \sin^{2j}(x)\cos^{2j}(x)
    \left(
        \sin^{2d-4j}(x)+\cos^{2d-4j}(x)
    \right)
    p^{d-j}(1-p)^j.
\end{align}
The normalization is
\begin{equation}
    \mathcal{N}
    =
    \left((1-p)\sin^2 x+p\cos^2 x\right)^d
    +
    \left((1-p)\cos^2 x+p\sin^2 x\right)^d,
\end{equation}
so that
\begin{equation}
    \tilde{q}_j=\frac{q_j}{\mathcal{N}},
    \qquad
    \tilde{q}'_j=\frac{q'_j}{\mathcal{N}}.
\end{equation}
The corresponding rotation angles are defined by
\begin{equation}
    \tan(\phi_j)
    =
    (-1)^{m-j}\tan^{d-2j}(x),
    \qquad
    \phi'_j=\phi_j+\frac{\pi}{2}.
\end{equation}

Rather than expanding in the physical angle $x$, which need not be small as $d$ grows, we instead express all angles in terms of the logical angle $\phi_0$. Let
\begin{equation}
    \alpha := |\tan(\phi_0)|.
\end{equation}
Then
\begin{equation}
    \tan(\phi_j)
    =
    (-1)^j \operatorname{sgn}(\tan\phi_0)\,
    \alpha^{1-2j/d},
\end{equation}
and hence
\begin{equation}
    \phi_j
    =
    \arctan\!\left(
        (-1)^j \operatorname{sgn}(\tan\phi_0)\,
        \alpha^{1-2j/d}
    \right),
    \qquad
    \phi'_j=\phi_j+\frac{\pi}{2}.
\end{equation}
This parameterization avoids assuming that $\tan(x)$ is small.

We can use the Mixing Lemma (Eq.~(\ref{Eq:Mixing})) and the associated bound in Eq.~(\ref{Eq:MixingVarBound}) to compute the error in approximating the mixture of rotations by a single rotation by the mean angle.

In the limit $dp \ll 1$, $\phi_0 \ll 1$, $\phi_0 \gg p^{d/2}$, the variance can be upper bounded by
\begin{equation}
    dp \phi_0^{2(1-1/d)} + \binom{d}{2}p^2 \phi_0^{2(1-2/d)} + ...
\end{equation}

\begin{figure*}
    \centering
    \includegraphics[width=0.9\linewidth]{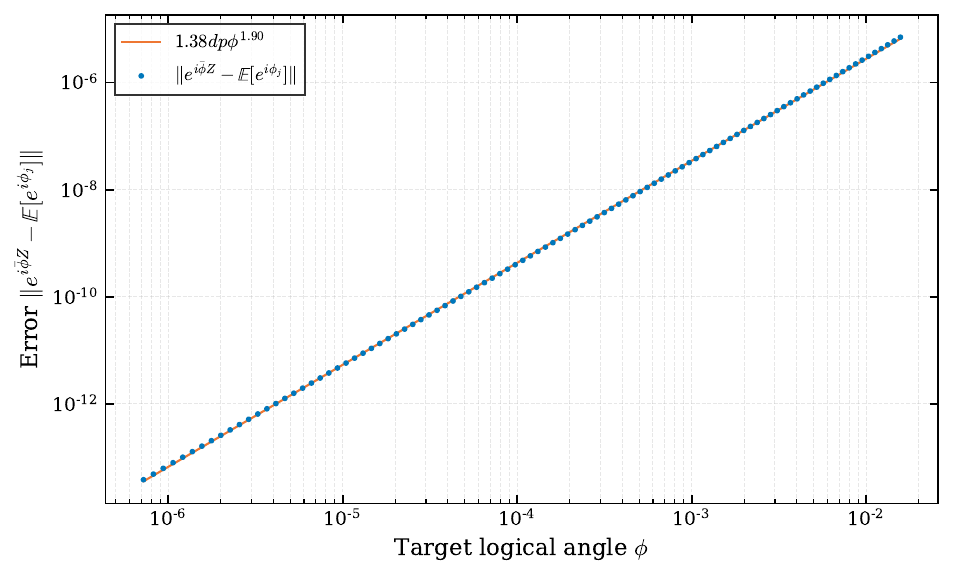}
    \caption{Computing the error using the Mixing Lemma, Eq.~(\ref{Eq:Mixing}). We consider a $d=13$ phase-flip repetition code, with a single-qubit phase flip error model on each qubit, where each single-qubit phase flip occurs with probability $p$. Here $p=10^{-3}$. We assume stabilizer measurement is noiseless. The resulting mixture of rotations can be computed analytically. We evaluate and plot $|| \exp(i\bar{\phi}Z) - \E[\exp(i\phi_jZ)]||$ for the corresponding distribution $\{q_j, \phi_j\}$. We observe that the error scaling is consistent with $\mathcal{O}(dp\phi^{(1-\frac{1}{d})})$. The best fit line is obtained through least squares on the relative error, with $f(\phi)=adp\phi^b$, and fitting parameters $a,b$.}
    \label{fig:MixingLemmaPlot}
\end{figure*}

This suggests that the Mixing Lemma can substantially improve the leading-order error scaling compared with directly sampling the mixture, without requiring the additional corrective rotations introduced in Refs.~\cite{toshio2025PartiallyFT,toshio2026starmutation}. We also provide numerical evidence for the Mixing Lemma approach, by considering the $m=1$ variant of the proposal in Ref.~\cite{toshio2025PartiallyFT} for a phase-flip repetition code, (equivalent to the proposal in Ref.~\cite{choi2023rotations}) subject to a simple single-qubit dephasing error model, and assuming noiseless syndrome extraction. The resulting mixture of rotations, including probabilities and angles, can be analytically computed and used to compute the error compared to implementing $\exp(i\bar{\phi}Z)$. The results are shown for $d=13$, $p=10^{-3}$ in Fig.~\ref{fig:MixingLemmaPlot}, and are consistent with the analysis above.

\subsection{A revised analysis of twirling, and decoherence by RUS teleportation}\label{App:Twirl}

Ref.~\cite{zhang2025low} discussed a twirled version of the RUS teleportation circuit, designed to eliminate the coherent term in the density matrix. However, we show through more careful analysis that it is the teleportation process, rather than the twirling method, that is decohering the errors. Moreover, the decoherence is only partial, giving an incoherent error only to leading order.

We will first recap the twirling analysis of Ref.~\cite{zhang2025low}. The density matrix of Eq.~(\ref{Eq:MixtureOfRotations}) can be re-expressed in the orthonormal basis $\ket{M(\phi)}, \ket{M(\phi)^\perp}$ where
\begin{equation}
    \ket{M(\phi)^\perp} = i\sin(\phi)\ket{+} + \cos(\phi)\ket{-} = Z_L \ket{M(\phi)}.
\end{equation}
Defining $\rho_\phi = \ket{M(\phi)}\bra{M(\phi)}$, the output density matrix is given by
\begin{equation}
    \tilde{\rho}_\phi = (\sum_{j = 0} q_j \cos^2(\phi_j - \phi)) \rho_\phi + \sum_{j \neq 0} q_j \bigg(\sin^2(\phi_j - \phi)Z_L\rho_\phi Z_L
    + \frac{i}{2}\sin(2(\phi_j -\phi)) ( Z_L\rho_\phi - \rho_\phi Z_L) \bigg).
\end{equation}
Each term has the form
\begin{equation}
    (1-a) \rho_\phi + a Z_L \rho_\phi Z_L + c Z_L \rho_\phi + c^* \rho_\phi Z_L.
\end{equation}

The magic state teleportation circuit for $R_z(\phi) \ket{\psi} $ is shown below, where the magic state is measured in the $Z$ basis. The teleportation implements $R_z(\phi)$ upon measuring $\ket{0}$ (probability 1/2) and implements $R_z(-\phi)$ upon measuring $\ket{1}$ --- necessitating a conditional (noisy) $R_z(2\phi)$ correction.
\begin{figure}[h]
    \centering
\begin{quantikz}
    \lstick{$\ket{\psi}$} & \ctrl{1} & & & \gate{R_z(2\phi)} & \\
    \lstick{$\ket{M(\phi)}$} & \targ{} & \meter{} &\setwiretype{c}  & \wire[u][1]{c} \\
\end{quantikz}\\
    \label{fig:TeleportationCircuit}
\end{figure}

Denote by $\mathcal{R}_z(\phi)[\cdot ] := R_z(\phi)(\cdot) R_z^\dagger(\phi) $. As noted in Ref.~\cite{zhang2025low}, we can write the action of the teleportation circuit as
\begin{equation}\label{Eq:UnTwirled}
    \mathcal{G}(\rho) = \Pi_{+} U \rho U^\dagger \Pi_{+} + \mathcal{R}_z(2\phi)[ \Pi_{-}  U \rho U^\dagger \Pi_{-}] 
\end{equation}
where $U$ is the CNOT gate, $\Pi_{\pm} = 0.5(I \pm Z)$ acts on the 2nd qubit, and $\mathcal{R}_z(2\phi)$ acts on the first qubit. We can also consider a twirled version of the circuit
\begin{equation}\label{Eq:Twirled}
    \mathcal{G'}(\rho) = \mathcal{R}_z(2\phi)[\Pi_{+} U (X_2 \rho X_2) U^\dagger \Pi_{+}] +  \Pi_{-}  U (X_2\rho X_2) U^\dagger \Pi_{-}.
\end{equation}
It can be easily verified that for an input state $\sigma \otimes \rho_\phi$, the output of both circuits is $\mathcal{R}_z(\phi) [\sigma] \otimes I$, when we assume $\mathcal{R}_z(2\phi)$ is implemented noiselessly. Similarly, for an input state $\sigma \otimes Z_L\rho_\phi Z_L$, the outputs of both circuits are $Z \mathcal{R}_z(\phi) [\sigma] Z \otimes I$. This is because $Z_L$ errors on the magic state are copied by the CNOT to the target state. As such, the diagonal elements of the density matrix are the same, whether we apply $\mathcal{G}$, or its twirled variant $\mathcal{G}'$.

However, the off-diagonal elements appear to be changed by applying the twirled variant:
\begin{align}
    \mathcal{G'}(\sigma \otimes Z\rho_\phi) &= \mathcal{R}_z(2\phi) [\Pi_{+} U X_2 \left(\sigma \otimes Z\rho_\phi \right) X_2 U^\dagger \Pi_{+}] +  \Pi_{-}  U X_2 \left(\sigma \otimes Z\rho_\phi \right) X_2U^\dagger \Pi_{-} \\
    &= - Z_1 Z_2  \mathcal{R}_z(2\phi) [\Pi_{+} U \left(\sigma \otimes X_2 \rho_\phi X_2 \right) U^\dagger \Pi_{+}] - Z_1 Z_2 \Pi_{-}  U  \left(\sigma \otimes X_2 \rho_\phi X_2 \right) U^\dagger \Pi_{-} \\
    &= - Z_1 \mathcal{R}_z(\phi) [\sigma] \otimes Z_2.
\end{align}
Note that this corrects a small error in Eq.~(S35) of Ref.~\cite{zhang2025low}. Similarly, the off-diagonal element for the regular teleportation channel $\mathcal{G}$ is:
\begin{align}\label{Eq:UnTwirledCoherent}
    \mathcal{G}(\sigma \otimes Z\rho_\phi) &= \Pi_{+}  U \left(\sigma \otimes Z\rho_\phi \right) U^\dagger \Pi_{+} + \mathcal{R}_{2\phi} [\Pi_{-} U  \left(\sigma \otimes Z\rho_\phi \right) U^\dagger \Pi_{-}]  \\
    &= Z_1 \mathcal{R}_\phi [\sigma] \otimes Z_2.
\end{align}
Observe that twirling appears to change the sign on the off-diagonal term. In Ref.~\cite{zhang2025low} this result is used to justify the claim that twirling can be used to remove the coherent error.\\

There are several issues with the above analysis. First, observe that the circuits in Eq.~(\ref{Eq:UnTwirled}) and Eq.~(\ref{Eq:Twirled}) are identical through propagation of the $X$ gate. This is expected for the twirling circuit without noise. However, note that the $X$ gate is applied to the noisy state $\rho$, so it cannot alter the noise already present in the state. For twirling to have an effect, it is necessary to prepare noisy approximations of $\rho_\phi$ or $\rho_{-\phi}$ that are affected by the same error channel (at least, the coherent error terms must be identical). However, observe from Eq.~(\ref{Eq:OrthonormalDecomp}) that the coherent term has prefactor $\sin(2(\phi_j - \phi))$. Under $\phi \rightarrow -\phi$, we also have $\phi_j \rightarrow -\phi_j$, and hence the sign on the coherent term flips. This violates the requirement of state-independent noise needed for successful twirling.

Second, observe that the output of passing the coherent term through the regular (not twirled) teleportation circuit is given by Eq.~(\ref{Eq:UnTwirledCoherent}). This expression assumes $\mathcal{R}_{2\phi}$ is implemented perfectly, which we will continue to assume for now. In this case, we simply want to reset the 2nd qubit to $\ket{0}$, because it is no longer required. The reset channel is $\ket{0}\bra{0} \mathrm{Tr}(\rho')$. We take the partial trace over the 2nd register. Because $\mathrm{Tr}(Z_2)=0$, this term vanishes. Thus, it is the act of teleportation which removes coherent error, rather than the twirling itself.  We have also confirmed this effect through numerical simulation of the teleportation channel.

However, note that the cancellation of coherent error (either in the unnecessary twirling analysis, or the equivalent untwirled case) only holds when $\mathcal{R}_{z}(2\phi)$ is implemented noiselessly. Unfortunately, this is in general not possible, and we must implement $\mathcal{R}_z(2\phi)$ through its own noisy RUS process. If the RUS cascade terminates in a rotation without error (e.g. an $S$ gate or fallback to synthesis with negligible synthesis error), then the coherent rotation is eliminated from the entire RUS cascade. If instead the RUS process is never terminated, the effect of coherent error can still be controlled---this occurs because the probability of needing trial $i$ decays exponentially with $i$, while the angles only double until they reach $\sim \pi/4$, after which they are reset using an $S$ gate and cease growing with $i$. This was the calculation from App.~\ref{App:MitigateCoherentErrors}; indeed, the calculation from the present appendix may be viewed as an alternative, more fine-grained viewpoint of how coherent errors are naturally mitigated in RUS.

\end{document}